\documentclass[sigconf]{acmart}

\AtBeginDocument{%
  }

\setcopyright{rightsretained}
\acmDOI{}
\acmConference[CHI '25 Workshop on Envisioning the Future of Interactive Health]{}{April 27th, 2025}{Yokohama, Japan}
\acmISBN{}
\copyrightyear{2025}
\acmYear{2025}

\usepackage{enumitem}

\begin{document}

\title[Over-Relying on Reliance]{Over-Relying on Reliance: Towards Realistic Evaluations of AI-Based Clinical Decision Support}

\author{Venkatesh Sivaraman}
\authornote{Equal contribution.}
\email{venkats@cmu.edu}
\orcid{0000-0002-6965-3961}
\author{Katelyn Morrison}
\authornotemark[1]
\email{kcmorris@andrew.cmu.edu}
\orcid{0000-0002-2644-4422}
\author{Will Epperson}
\orcid{0000-0002-2745-4315}
\authornotemark[1]
\email{willepp@cmu.edu}
\author{Adam Perer}
\email{adamperer@cmu.edu}
\orcid{0000-0002-8369-3847}
\affiliation{%
  \institution{Carnegie Mellon University}
  \city{Pittsburgh}
  \state{Pennsylvania}
  \country{USA}
}



\begin{abstract}
As AI-based clinical decision support (AI-CDS) is introduced in more and more aspects of healthcare services, HCI research plays an increasingly important role in designing for complementarity between AI and clinicians.
However, current evaluations of AI-CDS often fail to capture when AI is and is not useful to clinicians.
This position paper reflects on our work and influential AI-CDS literature to advocate for moving beyond evaluation metrics like Trust, Reliance, Acceptance, and Performance on the AI's task (what we term the ``trap'' of human-AI collaboration).
Although these metrics can be meaningful in some simple scenarios, we argue that optimizing for them ignores important ways that AI falls short of clinical benefit, as well as ways that clinicians successfully use AI.
As the fields of HCI and AI in healthcare develop new ways to design and evaluate CDS tools, we call on the community to prioritize ecologically valid, domain-appropriate study setups that measure the emergent forms of value that AI can bring to healthcare professionals. 
\end{abstract}





\maketitle

\section{Introduction}

\begin{figure*}
    \centering
    \includegraphics[width=\textwidth]{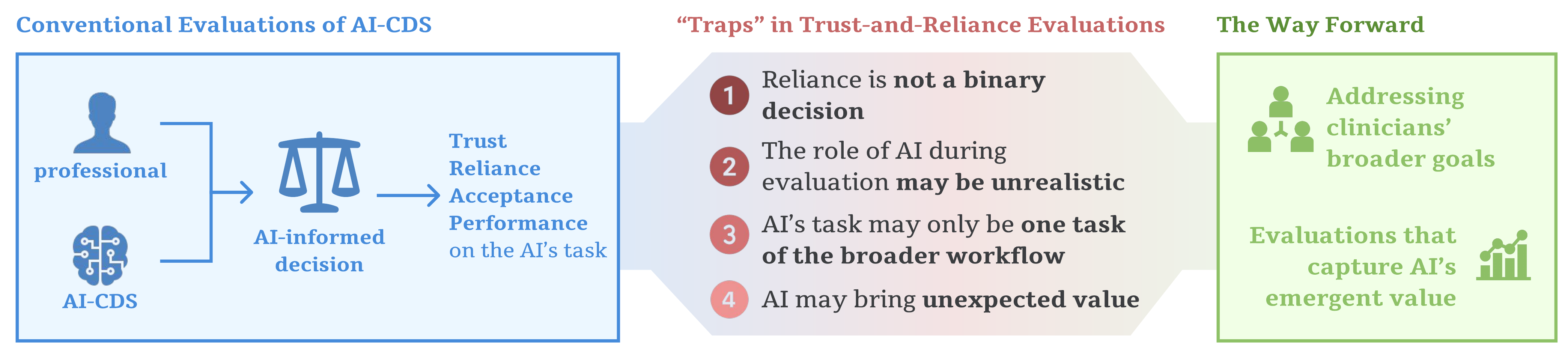}
    \caption{
    Many prior studies measure AI's success in clinical contexts in terms of their trust, reliance, acceptance, or performance on a task aligned with the AI's training. However, this fails to capture other forms of value that AI can provide to clinicians.} 
    \label{fig:trap}
\end{figure*}

AI-based clinical decision support (AI-CDS) systems aim to save clinicians time and effort while improving their overall accuracy and consistency.
Yet, many AI-CDS designs---such as those that offer a second opinion during decision-making---tend to fall short of these objectives outside controlled, laboratory settings~\cite{Beede2020,Sivaraman2023}.
HCI research, therefore, plays an essential role in designing AI-CDS that healthcare professionals can effectively use in practice.
Ideally, such systems enable \textit{complementarity}, in which the human-AI team achieves better outcomes than either the clinician or the AI alone~\cite{complimentarityBansal2021}.
However, despite significant empirical efforts, the field has yet to reach a consensus on which design strategies most effectively promote this collaborative benefit and deliver true clinical value~\cite{lai_towards_2023,yildirim_sketching_2024,vaccaro2024combinations}.
A case in point is explainable AI (XAI), which is no longer a primary focus of healthcare AI research despite being originally proposed by AI and HCI researchers to help clinicians calibrate their reliance on AI outputs. 
Does XAI have untapped potential that has yet to be fully realized, or is it addressing challenges that clinicians do not consider pressing? 
More broadly, as healthcare models continue to advance, how can we design systems and evaluations that genuinely align with the evolving needs of healthcare professionals?

We argue that the observed shortcomings of XAI in healthcare are one manifestation of a bigger limitation in current HCI approaches to evaluating decision support:
a predominant focus on measuring \textbf{Trust}, \textbf{Reliance}, \textbf{Acceptance}, and \textbf{Performance} in human-AI teams (collectively, TRAP).
Each of these metrics represents a similar conception of human-AI collaboration, in which the user either accepts or rejects an AI output to make a decision that is directly aligned with the AI's task (\textit{e.g.}, a binary diagnosis). 
This setup has influenced a great deal of research on AI-CDS, including our own work~\cite{sivaraman2025static,Sivaraman2023,turri2024transparency}, influential studies in clinical journals~\cite{Tschandl2020,Jacobs2021}, and new studies examining reliance on generative AI~\cite{goh_large_2024}. 
But due to the pitfalls described below, evaluation studies using these metrics may well be falling into a ``trap'' that must be addressed if we are to benchmark our progress toward effective AI-CDS.

\section{Locating the Traps in Trust-and-Reliance Evaluations}

At face value, there may appear to be good reasons to define successful AI-CDS in terms of appropriate trust and reliance.
Lee and See's foundational review 
of trust in automation draws a direct link between appropriate trust---willingness to rely on a system under uncertainty or vulnerability---and mitigation of high-stakes automation failures~\cite{lee_trust_2004}.
When opaque, unpredictable, or error-prone systems perform critical tasks on which people's survival depends, calibrating users' trust and reliance is undoubtedly essential.

However, this framing assumes an AI model predicts the exact same task that a human performs.
Modern AI systems can be integrated into healthcare workflows in a much wider variety of ways, such as retrieving information, forecasting possible outcomes, coordinating a multidisciplinary team, communicating with patients, and generating ideas for inspiration. 
Yet trust, reliance, acceptance, and team performance still serve as the dominant proxy metrics for successful AI-CDS.
As a result, unforeseen barriers to adoption arise in more open-ended evaluations that cannot be explained through the lens of reliance~\cite{Beede2020,Sivaraman2023}.
Below we describe four pitfalls in evaluations of AI-CDS that can cause the reliance-based paradigm to break down, potentially preventing us from measuring the true value of AI in clinical decision making contexts:



\begin{enumerate}[leftmargin=*]
    \item \textbf{Reliance is Not a Binary Decision.} 
    Evaluations of appropriate reliance on AI-CDS often require that users either fully accept or fully reject a recommendation~\cite{Jussupow2020}.
    However, real-world decision making tasks are often much more complex than a single accept/reject decision.
    The ability to apply discretion and bring in prior experience and expertise that is inaccessible to the AI is a key reason humans retain decision-making power. 
    Measuring reliance as participants' binary agreement with an AI recommendation, as is currently done even for complex systems such as LLMs~\cite{kim_uncertainty_2024}, fails to capture when participants \textit{partially} rely on AI outputs.
    For example, users may be influenced or inspired by AI-generated content, or they may selectively choose aspects of an AI recommendation to follow based on the urgency of a case~\cite{Sivaraman2023}.
    These forms of reliance do not fit the dichotomous structure of acceptance or rejection, yet they represent important ways AI can be useful.
    

\item \textbf{The Role of the AI During Evaluation May Be Unrealistic.}
Studies on AI-assisted clinical decision-making typically use a sequential decision-making setup in which the clinician reviews a case, optionally provides an initial assessment, then makes a decision with the use of an AI prediction~\cite{Sivaraman2023,Tschandl2020,prinster_care_2024}.
However, at the point of deployment, AI-CDS tools may take on different roles in the clinical workflow, such as bedside alerts or triage systems~\cite{yildirim_sketching_2024}.
Whereas clinicians may give credence to an AI in a standardized evaluation study, its role in a deployed workflow may render it superfluous (\textit{e.g.}, if shown after inputting an order).
Conversely, clinicians may find utility in AI-generated information whose apparent purpose is not explicitly decision support (for example, creating presentation materials~\cite{Yang2019}).
These failures and potential successes go unanticipated if AI outputs are only evaluated in a standard reliance setup.

\item \textbf{The AI's Task May Be Just One Part of the Human's Goal.}
Most AI-CDS evaluations simplify the user's workflow into just the aspects related to the AI's task (\textit{e.g.}, predicting risk of mortality in one year or detecting a specific finding in an X-ray).
Decision-making tasks are often conceptualized retrospectively in this format, even if the algorithm was actually used differently in practice~\cite{adams_prospective_2022}.
While this setup makes the measurement of trust-and-reliance metrics more tractable, it crucially removes complexity that has a direct impact on how the human might (or might not) use the AI.
For example, studies showing that incorrect AI diagnoses can ``mislead'' clinicians~\cite{Tschandl2020,Jacobs2021} disregard the reality that such advice could still prompt clinicians to take other actions (\textit{e.g.}, running tests or scheduling a prompt follow-up) that might ultimately improve outcomes~\cite{Sivaraman2023}.
Therefore, when the AI performs just a small subset of the tasks involved in achieving the user's goals, people's reliance on the AI when the other tasks are stripped away is unlikely to reflect true usage.

\item \textbf{The AI May Bring Value in Unexpected Ways.}
By focusing on the AI's effect on decision task performance---which often falls short of expectations---we may lose sight of opportunities for the AI to support other user goals beyond ``making a decision.'' 
We can view these alternative uses for AI as \textit{appropriations}, or unanticipated transformations of the existing affordances of a system~\cite{zhang_resilience_2023}.
AI explanations, for example, can be appropriated to facilitate communication between care providers and patients or justify decisions to colleagues~\cite{corti2024moving,sivaraman2025static}.
Early-stage design work has been instrumental in identifying these emergent use cases~\cite{corti2024moving,turri2024transparency},
but they are more challenging and just as important to explore in working AI systems.
By grounding feedback in real AI behaviors, exploratory human-centered evaluations with working systems can help refocus evaluations toward these sources of value.
\end{enumerate}

\section{The Way Forward: Identifying and Measuring AI's Value to Clinicians}

We call on the growing HCI$+$Health community to develop and disseminate \textbf{evaluation strategies to match the real forms of value that AI can bring healthcare providers}. 
Most importantly, researchers conducting quantitative evaluations of AI-CDS should decide on the behaviors they want to measure, then create the most ecologically valid experimental setup possible to measure that behavior.
For example, systems meant to improve diagnostic performance in radiology could be evaluated by asking clinicians to provide their diagnosis as unstructured text (as they would for a real patient) rather than simply measuring their acceptance of a diagnosis, which is typically only a minor part of a radiologist's workflow~\cite{calisto2023assertiveness,prinster_care_2024}.
Simulating and evaluating physician-patient and care team interactions could be another promising way to measure how AI supports communication and collaboration.

Despite the implications of the name ``TRAP,'' we do not suggest that trust and reliance metrics should be abandoned entirely in evaluations of CDS.
Rather, when appropriate, they can be made more realistic by \textbf{constructing experiments to situate AI usage within the clinician's broader workflow}.
For example, studies can measure how AI supports clinicians' downstream tasks and end goals, rather than their performance on the AI's task.
Similarly, while explanations' benefits to trust and reliance may be unclear, they could still be a valuable addition to AI-CDS if their interfaces and associated evaluation metrics are designed around clinicians' communication needs.

Importantly, when synthesizing findings across AI-CDS evaluations, we believe that \textbf{the results of previous evaluations should not be expected to generalize across different \textit{roles} that AI could play in a healthcare workflow}. 
An AI system may act as the user's assistant in one context~\cite{goh_large_2024}, their expert colleague in another~\cite{Sivaraman2023}, or their reference manual or alert system~\cite{adams_prospective_2022}.
Each possible role provides different benefits and value to clinicians, translating to different desired metrics for evaluating the human-AI team's success. 
Results from one may not transfer to another despite the use of similar TRAP-based metrics, leading to confusion in the literature about when human-AI collaboration does and does not work~\cite{vaccaro2024combinations}.
Future literature could delineate CDS systems by the type of insights they provide (for example, descriptive, predictive, or prescriptive), the clinician population they are designed for, or the degree to which they transform existing workflows.

We believe that \textbf{interdisciplinary collaborations and open-ended qualitative research with clinicians will play a critical role in determining what roles AI should play in clinicians' work, as well as how to design those AI systems to promote those valued roles and behaviors}. 
This requires involving clinicians as core collaborators in HCI+Health research studies, not only during early design stages and summative evaluations but \textit{throughout} the development process.
In particular, we have found it helpful to conduct exploratory usability evaluations using functional models trained on real data, which can help identify both misalignments and opportunities to create value~\cite{sivaraman2025static,Sivaraman2023}.
For example, after our exploratory study of XAI with ICU clinicians showed that AI-based treatment recommendations were largely misaligned with intensive care clinicians' expectations~\cite{Sivaraman2023}, our design efforts shifted towards more valuable roles AI could play in their decision-making.
By adopting a broader space of both methods and metrics to evaluate AI-CDS, the fields of HCI and AI in healthcare have an opportunity to cultivate research across disciplines that fulfills the long-held dream of supporting healthcare work with AI.
\begin{acks}
Thanks to John Zimmerman, Dominik Moritz, and Lingwei Cheng for conceptual feedback on these ideas.
\end{acks}

\bibliographystyle{ACM-Reference-Format}
\bibliography{references}


\end{document}